\documentclass[intlimits,twoside,a4paper]{article}
\usepackage{amsmath,amssymb}
\usepackage{graphicx}
\usepackage{dcolumn}
\usepackage{bm}

\usepackage[T2A]{fontenc}
\usepackage[cp1251]{inputenc}
\usepackage{cmpj}

\issue{2011}{14}{2}{23703}

\doinumber{10.5488/CMP.14.23703}

\title[Theoretical studies of the spin Hamiltonian parameters]%
{Theoretical studies of the spin Hamiltonian parameters for the
two tetragonal {Cu}$^{2+}$ centers in the calcined catalysts
CuO--ZnO}

\author[H.M.~Zhang, S.Y. Wu, Z.H. Zhang]
{H.M.~Zhang\refaddr{a1,a2}\thanks{E-mail:
huamingzhang66@gmail.com}\,\,, S.Y.~Wu\refaddr{a1,a3},
Z.H.~Zhang\refaddr{a1}}
\addresses{\addr{a1} Department of Applied Physics, University of
Electronic Science and Technology of China, \\610054~Chengdu, China
\addr{a2} Key Laboratory of Nondestructive Test, Ministry of
Education, Nanchang Hangkong University, 330063~Nanchang, China
\addr{a3} International Centre for Materials Physics, Chinese
Academy of Science, 110016~Shenyang, China}
\authorcopyright{H.M.~Zhang, S.Y. Wu, Z.H. Zhang, 2011}
\date{Received October 31, 2010, in final form April 15, 2011}

\begin{document}

\maketitle

\begin{abstract}
The spin Hamiltonian parameters for the two Cu$^{2+}$ centers $A1$
and $A2$ in the calcined catalysts CuO--ZnO are theoretically
investigated using the high order perturbation formulas of these
parameters for a $3d^{9}$ ion in tetragonally elongated octahedra.
In the above formulas, the tetragonal field parameters $Ds$ and
$Dt$ are determined from the superposition model, by considering
the relative axial elongation of the oxygen octahedron around the
{Cu}$^{2+}$ due to the Jahn-Teller effect. Based on the
calculations, the relative elongation ratios of about $5\%$ and
$3\%$ are obtained for the tetragonal {Cu}$^{2+}$ centers $A1$ and
$A2$, respectively. The theoretical spin Hamiltonian parameters
are in good agreement with the observed values for both systems.
The larger axial elongation in center $A1$ is ascribed to the more
significant low symmetrical (tetragonal) distortion of the
Jahn-Teller effect. The local structures characterized by the
above axial elongations are discussed.
\pacs 76.30.Fc, 75.10.Dg, 71.70.Ch
\end{abstract}

\section{Introduction}

\quad The calcined catalysts CuO--ZnO are the systems widely
applied in the fields of CO oxidation~\cite{1,2,3,4,5,6}. The
catalytic properties can be closely related to the electronic
states and local structures of Cu in ZnO, which may be
conveniently investigated with the aid of electron paramagnetic
resonance (EPR) technique. {Cu}$^{2+}$ $(3d^{9})$ is usually
treated as a model system with one equivalent $3d$ hole, having a
ground state and a single excited state under ideal octahedral
crystal-fields~\cite{7}. In CuO--ZnO catalysis, two non-equivalent
Cu$^{2+}$ locate on slightly tetragonally (D$_{4\mathrm{h}}$) elongated
octahedra, with four coplanar {Cu}$^{2+}$--{O}$^{2-}$ bond lengths
shorter than the axial ones~\cite{8,9,10,11}. The properties of
the above structure may play an important role in the
catalytic behaviour of these systems. That is why their local
structure features are worth being further investigated. For
example, the EPR experiments were carried out for the calcined
catalysts CuO--ZnO, and the spin Hamiltonian parameters (the
anisotropic $g$ factors $g_{\parallel}$ and $g_{\perp}$ and the
hyperfine structure constants $A_{\parallel}$ and $A_{\perp}$)
were also measured for the two signals $A1$ and $A2$~\cite{12}.
The signals $A1$ and $A2$ are attributable to isolated Cu$^{2+}$
ions automically dispersed in the lattice of ZnO and located in
tetragonally distorted octahedral cavities~\cite{12}. The observed
EPR spectra reveal positive $g$ anisotropy $\Delta g$
($=g_{\parallel} - g_{\perp}$) and a much larger $A_{\parallel}$
in magnitude than $A_{\perp}$, characteristic of a $3d^{9}$ ion
under a tetragonally elongated octahedron. This point is quite
similar to the case of {Cu}$^{2+}$/{Al}$_{2}${O}$_{3}$
systems~\cite{13}. According to references~\cite{12,14}, the
difference between signals $A1$ and $A2$ may be ascribed to the
two slightly different types of Cu$^{2+}$ in different positions
and unlike local environments in the CuO--ZnO systems.
Specifically, the higher $g_{\parallel}$ and lower $A_{\parallel}$
values for center $A2$ can originate from the samples suffering
stronger axial interaction~\cite{14,12}. However, the above
experimental results have not been theoretically explained so far,
and the data on the local structures of the Cu$^{2+}$ centers have
not been obtained yet. Since the analysis of the EPR spectra can
provide useful data on the electronic states and local structures
for Cu$^{2+}$ in the calcined catalysts CuO--ZnO, which would be
helpful in understanding the properties of the catalysts, further
studies on the above EPR experimental results are of particular
scientific and practical significance. In this work, the spin
Hamiltonian parameters are theoretically investigated for the two
Cu$^{2+}$ centers $A1$ and $A2$ in the calcined catalysts CuO--ZnO
using the high order perturbation formulas of these parameters for
a $3d^{9}$ ion in tetragonally elongated octahedra. In the above
formulas, the tetragonal field parameters $Ds$ and $Dt$ are
determined from the superposition model, by considering the
relative axial elongation (in terms of the relative elongation
ratio $\rho$ of the axial {Cu}$^{2+}$--{O}$^{2-}$ bond lengths
related to the average bond length) of the oxygen octahedron
around the Cu$^{2+}$ due to the Jahn-Teller effect. The
calculation results and the local structures of the Cu$^{2+}$
centers are discussed.

\section{Calculation}

In the calcined catalysts CuO--ZnO, the original Cu$^{2+}$ in the
form of CuO may be located in the center of oxygen octahedra. As a
Jahn-Teller ion, Cu$^{2+}$ will suffer the Jahn-Teller effect via
stretching the two {Cu}$^{2+}$--{O}$^{2-}$ bonds along the
$\emph{C}_{4}$ axis and contracting the other four bonds in the
perpendicular plane, which reduces the local symmetry to
tetragonal (an elongated octahedron). This point has been
illustrated in various studies on Cu$^{2+}$ in oxygen
octahedra~\cite{15}. Thus, the local structures of the Cu$^{2+}$
centers can be characterized by the relative tetragonal elongation
ratio (labeled as $\rho$) in the [{CuO}$_{6}$]$^{10-}$ clusters.
For a {Cu}$^{2+}$ $(3d^{9})$ ion in tetragonally elongated
octahedra, the lower $^{2}E_{\rm g}$ irreducible representation
may be separated into two orbital singlets $^{2}B_{1 \rm g}$ and
$^{2}A_{1 \rm g}$\,, with the former lying lowest~\cite{7,15}.
Meanwhile, the upper $^{2}T_{2 \rm g}$ representation would split
into an orbital singlet $^{2}B_{2 \rm g}$ and a doublet
$^{2}E_{\rm g}$~\cite{7,15}. The perturbation formulas of the spin
Hamiltonian parameters for a tetragonally elongated $3d^{9}$
cluster can be expressed as~\cite{16}:
\begin{eqnarray}\label{1}
g_{\parallel}&=&g_{\rm s}+8{k}\zeta_{\rm d}/E_{1}+k\zeta_{\rm d}^{2}/E^{2}_{2}
+4k\zeta_{\rm d}^{2}/(E_{1}E_{2})
+g_{\rm s}\zeta_{\rm d}^{2}\left[1/E^{2}_{1}-1/\left(2E_{2}^{2}\right)\right]\nonumber\\
&&+k\zeta^{3}_{\rm d}\left[4/\left(E_{1}E^{2}_{2}\right)-1/E^{3}_{2}\right]
-2k\zeta^{3}_{\rm d}\left[2/\left(E^{2}_{1}E_{2}\right)
-1/\left(E_{1}E^{2}_{2}\right)\right],\nonumber\\
g_{\perp}&=&g_{\rm s}+2k\zeta_{\rm d}/E_{2}+k\zeta^{2}_{\rm
d}\left[2/\left(E_{1}E_{2}\right)
-1/E^{2}_{2}-4/(E_{1}E_{2})\right]\nonumber\\
&&+2g_{\rm s}\zeta^{2}_{\rm d}/E^{2}_{1}+k\zeta^{3}_{\rm
d}\left(4/E^{2}_{1}
-1/E^{2}_{2}\right)/(2E_{2}),\nonumber\\
A_{\parallel}&=&\emph{P}\left[\left(-\kappa-4\emph{N}/7\right)
+(g_{\parallel}-g_{\rm s})+3(g_{\perp}-g_{\rm s})/7\right],\nonumber\\
A_{\perp}&=&\emph{P}\left[\left(-\kappa+2\emph{N}/7\right)-(g_{\perp}-g_{\rm
s})/7\right]
\end{eqnarray}
where $g_{\rm s}\approx 2.0023$ is the spin-only value. Parameter
k is the orbital reduction factor, which is equivalent to the
covalency factor \emph{N}. Parameter $\kappa$ is the core
polarization constant. $\zeta_{\rm d}$ and \emph{P} are,
respectively, the spin-orbit coupling coefficient and the dipolar
hyperfine structure parameter for the $3d^{9}$ ion in crystals.
They can be written in terms of the corresponding free-ion values,
i.e., $\zeta_{\rm d}\approx\emph{N}\zeta^{0}_{\rm d}$ and
$\emph{P}\approx\emph{N}\emph{P}_{0}$\,. $E_{1}$ and $E_{2}$ are
the energy separations between the excited $^{2}B_{2 \rm g}$ and
$^{2}E_{\rm g}$ and the ground $^{2}B_{1 \rm g}$ states~\cite{16}:
\begin{eqnarray}\label{2}
E_{1}&=&10Dq,\nonumber\\
E_{2}&=&10Dq-3Ds+5Dt.
\end{eqnarray}
Here $Dq$ is the cubic field parameter, and $Ds$ and $Dt$ are the
tetragonal field parameters.

For the Jahn-Teller elongated $[\text{CuO}_{6}]^{10-}$ clusters,
the parallel and perpendicular bond lengths can be expressed in
terms of the relative tetragonal elongation ratio $\rho$ and the
reference distance $\emph{R}$ as:
$\emph{R}_{\parallel}\approx\emph{R}(1 + 2\rho)$ and
$\emph{R}_{\perp}\approx\emph{R}(1-\rho)$. Thus, the cubic and
tetragonal field parameters are determined from the superposition
model~\cite{16} and the geometrical relationship of the Cu$^{2+}$
centers:
\begin{eqnarray}\label{3}
Dq &\approx& (3/4)\bar{A}_{4}\left(1-\rho\right)^{-t_{4}},\nonumber\\
Ds &\approx& (2/7)\bar{A}_{2}\left[(1-\rho)^{-t_{2}}-(1+2\rho)^{-t_{2}}\right],\nonumber\\
Dt &\approx&
(16/21)\bar{A}_{4}\left[(1-\rho)^{-t_{4}}-(1+2\rho)^{-t_{4}}\right].
\end{eqnarray}
Here $t_{2}\approx 3$ and $t_{4}\approx 5$ are the power-law
exponents in view of the ionic nature of the bonds~\cite{17}.
$\bar{A}_{2}$ and $\bar{A}_{4}$ are the rank-2 and rank-4
intrinsic parameters, respectively. For octahedral $3d^{n}$
clusters, the relationships $\bar{A}_{4}\approx (3/4)Dq$ and
$\bar{A}_{2}\approx 10.8\bar{A}_{4}$~\cite{17,18,19} are proved
valid in many crystals and are reasonably applied here. Thus, the
$g$ factors, especially the anisotropy $\Delta g (=
g_{\parallel}-g_{\perp})$ is connected with the tetragonal field
parameters and hence with the local structure (i.e., the relative
tetragonal elongation ratio $\rho$) of the systems studied.

According to the optical spectra for Cu$^{2+}$ in some
oxides~\cite{20,21}, the values $\bar{A}_{4}\approx 800$~cm$^{-1}$
and $\emph{N}\approx0.82$ can be obtained. The spin-orbit coupling
coefficient $\zeta_{\rm d}$ and the dipolar hyperfine structure
parameter are acquired for the  systems studied using the free-ion
data $\zeta^{0}_{\rm d}\approx829$~cm$^{-1}$~\cite{22} and
$\emph{P}_{0}\approx416\times10^{-4}$~cm$^{-1}$~\cite{23}. The
core polarization constant is taken as the expectation value
$0.3$~\cite{22} for $3d^{n}$ ions in crystals. Thus, only the
relative tetragonal elongation ratio $\rho$ is unknown in the
formulas of the spin Hamiltonian parameters. Substituting these
values into equation~(\ref{1}) and fitting the calculated results to the
experimental data, one can obtain
\begin{equation}\label{4}
\rho\approx 5\% \quad \text{and} \quad 3\%
\end{equation}
for the two centers $A1$ and $A2$, respectively. The corresponding
theoretical results are shown in table~\ref{tab1}.
\begin{table}[htcp]
\caption{ The $g$ factors and the hyperfine structure constants
(in $10^{-4}$~{cm}$^{-1}$) for the two tetragonal Cu$^{2+}$
centers in the calcined catalysts CuO--ZnO.}\label{tab1}%
\vspace{1ex}
\begin{center}
  \begin{tabular}{|l|c|c|c|c|c|}
\hline
  & &$g_{\parallel}$&$g_{\perp}$&$A_{\parallel}$&$A_{\perp}$\\\cline{1-6}
  $A1$     &Cal.&$2.329$ & $2.070$ & $-142.5$ &$18.8$\\\cline{2-6}
           &Expt. [12]&$2.325$ & $2.075$ & $-140.1$ &$18.7$\\\cline{1-6}
  $A2$     &Cal.&$2.367$ & $2.083$ & $-139.4$ &$17.3$\\\cline{2-6}
           &Expt. [12]&$2.366$ & $2.085$ & $-130.7$ &$18.7$\\\hline
  \end{tabular}
\end{center}
\end{table}

\section{Discussion}

Table~\ref{tab1} reveals that the calculated spin Hamiltonian parameters
for the two centers $A1$ and $A2$ in the calcined catalysts
CuO--ZnO based on the relative tetragonal elongation ratios in
equation~(\ref{4}) are in good agreement with the observed values. Thus,
the experimental EPR spectra~\cite{12} for both systems are
satisfactorily interpreted in this work, and the local structure
information is obtained.

\begin{enumerate}
\item{} The EPR spectra for the Cu$^{2+}$ centers can be characterized
by the positive anisotropy $\Delta g$. The anisotropy largely
depends upon the tetragonal distortion (i.e., the tetragonal field
parameters $Ds$ and $Dt$) arising from the relative tetragonal
elongation ratio $\rho$ ($\approx 3\% -5\%$). Further, the
tetragonal elongations are of the Jahn-Teller nature via
relaxation of the parallel bond lengths and contraction of the
perpendicular ones. Similar tetragonal elongation of the oxygen
octahedra was also reported for Cu$^{2+}$ on the {Al}$^{3+}$ site
in {LaSrAlO}$_{4}$~\cite{15}. The relative elongation ratio $\rho$
in the $A1$ center larger than in the $A2$ center may be
attributed to the more significant Jahn-Teller distortion in the
former. Thus, one can expect that the two centers $A1$ and $A2$
may be distorted in different ways. The difference in the
Jahn-Teller distortion strength is likely due to the
[{CuO}$_{6}$]$^{10-}$ clusters at different octahedral cavities
and hence differ from local environments in the CuO--ZnO systems.
Similar difference in the Jahn-Teller distortion was also reported
for the two trigonal {Ti}$^{3+}$ (with the same spin ${S}=1/2$)
centers at non-equivalent octahedral {Al}$^{3+}$ sites in
{LaMgAl}$_{11}${O}$_{19}$ corresponding to two different sets of
EPR spectra~\cite{24}.
\item{}The hyperfine structure constants for the Cu$^{2+}$ centers are
close to each other, which can be illustrated by the similar
spectral parameters and the core polarization constant. However,
the parallel component $(A_{\parallel})$ is smaller in $A2$ center
than that in $A1$ center. From equation~(\ref{1}) it is seen that
$A_{\parallel}$ is effected by $g_{\parallel}-g_{\rm s}$\,, while
$A_{\perp}$ remains almost the same because $g_{\perp}$ shows the
similar values for both centers. Thus, the larger $g_{\parallel}$
in $A2$ center may somewhat cancel the negative isotropic term
(proportional to the core polarization constant) and then leads to
the lower magnitude of $A_{\parallel}$\,. The signs of the
experimental hyperfine structure constants were not given
in~\cite{12}. Nevertheless, the theoretical studies in this work
indicated negative signs of hyperfine structure constants
$A_{\parallel}$\,.
\item{} The studied Cu$^{2+}$ centers in the calcined catalysts
CuO--ZnO are ascribed to the octahedral [{CuO}$_{6}$]$^{10-}$
clusters embedded in the ZnO host. Although ZnO exhibits some
covalency, the Cu$^{2+}$ centers are mainly ionic, characterized
by the covalency factor $N$ ($\approx0.82$). In addition, the
spin-orbit coupling coefficient ($\approx151$~cm$^{-1}$)~\cite{25}
of the ligand oxygen is much smaller than that ($\approx
\nolinebreak829$~cm$^{-1}$)~\cite{18} of the central ion
Cu$^{2+}$. Thus, contributions to the spin Hamiltonian parameters
from the ligand orbital and spin-orbit coupling interactions can
be regarded as very small and negligible for simplicity.
\item{}
The calculation errors of the present work can be discussed as
follows. First, the approximation of the theoretical model (i.e.,
crystal-field model) and formulas may lead to some errors. Second,
the spectral parameters $Dq$ (or $\bar{A}_{4}$) and $\emph{N}$
obtained from those for Cu$^{2+}$ in some oxides can effect the
final results. As $Dq$ varies by $10\%$\,, the relative
elongations ratio $\rho$ and the spin Hamiltonian parameters would
be modified by about $0.5\%$\,, because the tetragonal distortion
(or~$\rho$) and the resultant $\Delta g$ [see equation~(\ref{3})] have a
little relation to $Dq$. While ${N}$ changes by $10\%$\,, the
final $\rho$ and spin Hamiltonian parameters deviate by only
$0.4\%$\,, suggesting that covalency affects mainly the average of
the $g$ factors and brings forward a few effects on $\rho$ and
$\Delta g$. Third, the errors can also be introduced from the
approximation of the relationship $\bar{A}_{2} \approx
10.8\bar{A}_{4}$~\cite{17,18,19}, which would slightly modify the
tetragonal field parameters. The errors are estimated to be no
more than $1\%$ for $\rho$ and the spin Hamiltonian parameters
when the ratio $\bar{A}_{2}/\bar{A}_{4}$ varies within the widely
accepted range of 9--12. Finally, the uncertainty of the core
polarization constant $\kappa$ used in the present calculations of
the hyperfine structure constants may lead to some errors. Since
this value is the expectation result for transition-metal ions in
crystals, the effects of the above errors on the hyperfine
structure constants may be regarded as very small or negligible.
\end{enumerate}

\section{Summary}

\quad The spin Hamiltonian parameters and the local structures of
the tetragonal Cu$^{2+}$ centers $A1$ and $A2$ in the calcined
catalysts CuO--ZnO are theoretically investigated from the
perturbation formulas for a $3d^{9}$ ion in tetragonally elongated
octahedra. The oxygen octahedra around Cu$^{2+}$ are found to
suffer the relative tetragonal elongation ratios of about $3\%$
and $5\%$ due to the Jahn-Teller effect for centers $A1$ and $A2$,
respectively. The axial elongation for center $A1$ larger than for
center $A2$ can be attributed to the more significant tetragonal
Jahn-Teller distortion
in the former.

\section*{Acknowledgements}
\quad This work was financially supported by the Support Program for
Academic Excellence of UESTC.

\newpage
\ukrainianpart

\title[]%
{Теоретичні дослідження параметрів спінового гамільтоніану для двох тетрагональних центрів Cu$^{2+}$  в кальцинованих каталізаторах CuO--ZnO}

\author[Х.M.~Жанг, С.Й. Ву, Ж.Х. Жанг]
{Г.M.~Жанг\refaddr{a1,a2}, С.Й.~Ву\refaddr{a1,a3},
Ж.Г.~Жанг\refaddr{a1}}

\addresses{\addr{a1} Університет електроніки і технології Китаю, 610054~Ченду, Китай
\addr{a2} Провідна лабораторія неруйнівних випробувань, Міністерство освіти, Університет  Нанчанг Ганконг, 330063~Нанчанг, Китай
\addr{a3} Міжнародний центр фізики матеріалів, Китайська академія наук,  110016~Шеньянг, Китай}

\makeukrtitle

\begin{abstract}
Параметри спінового гамільтоніану для двох центрів
Cu$^{2+}$, $A1$ і $A2$, у кальцинованих каталізаторах CuO--ZnO досліджуються теоретично, використовуючи
формули теорії збурень високого порядку для цих параметрів для
іона  $3d^{9}$ у тетрагонально видовжених октаедрах. В цих формулах параметри тетрагонального поля  $Ds$ і
$Dt$ визначаються з суперпозиційної моделі, шляхом розгляду відносного аксійного видовження  октаедра кисню навколо
{Cu}$^{2+}$ відповідно до ефекту Яна-Теллера. Базуючись на цих розрахунках, отримано коефіцієнти відносного видовження  приблизно $5\%$ і
$3\%$ для тетрагональних  {Cu}$^{2+}$  центрів $A1$ і $A2$, відповідно. Параметри теоретичного спінового гамільтоніану доб\-ре
узгоджуються зі спостережуваними величинами  для обох систем. Більше аксійне видовження  в центрі  $A1$  приписується до вагомішої низько симетричної
(тетрагональної) дисторсії ефекту Яна-Телера. Обговорюються локальні структури, що  характеризуються  вище згаданим аксійним видовженням.
\keywords EPR, {Cu}$^{2+}$, CuO--ZnO, дефектні структури
\end{abstract}

\label{last@page}

\newpage
\begin{thebibliography}{99}
%
\bibitem{1} Kurr P., Kasatkin I., Girgsdies F., Trunschke~A., Schl\"{o}gl~R., Ressler~T.,
Appl. Catal. A, 2008, \textbf{348}, 153;
\doi{10.1016/j.apcata.2008.06.020}.
%
\bibitem{2} Bao J., Liu Z.L., Zhang Y., Tsubaki N., Catal. Commun., 2008, \textbf{9},
913;\\
\doi{10.1016/j.catcom.2007.09.034}.
%
\bibitem{3} Guo P.J., Chen L.F., Yu G.B., Zhu~Y., Qiao~M.H., Xu~H.L., Fan~K.N.,
Catal. Commun., 2009, \textbf{10}, 1252;
\doi{10.1016/j.catcom.2009.01.032}.
%
\bibitem{4} Liu L., Zhao T.S., Ma Q.X., Shen Y.F., J.~Natur. Gas Chem.,
 2009, \textbf{18}, 375;\\ \doi{10.1016/S1003-9953(08)60121-8}.
 %
\bibitem{5} Kang S.H., Bae~J.W., Prasad~P.S.S., Oh~J.H., Jun~K.W., Song S.L., Min K.S.,%
J.~Industrial Engin. Chem., 2009, \textbf{15}, 665;
\doi{10.1016/j.jiec.2009.09.041}.
%
\bibitem{6} Toyir J., Miloua R., Elkadri~N.E., Nawdali M., Toufik H., Miloua~F., Saito~M., %
Phys. Procedia, 2009, \textbf{2}, 1075; \doi{10.1016/j.phpro.2009.11.065}.%
%
\bibitem{7} Abragam A., Bleaney B., Electron Paramagnetic Resonance of Transition Ions.
Oxford University Press, London, 1970.
%
\bibitem{8} Porta P., Fierro G., Lo Jacono M., Moretti G.,
Catal. Today, 1988, \textbf{2}, 675; \\
\doi{10.1016/0920-5861(88)85031-4}.
%
\bibitem{9} Velu S., Suzuki K., Gopinath C.S., Yoshida~H., Hattori~T.,
Phys. Chem. Chem. Phys., 2002, \textbf{4}, 1990;
\doi{10.1039/b109766k}.
%
\bibitem{10} Porta P., Derossi S., Ferraris G., Jacono~M.L., Minelli~G., Moretti~G.,
J.~Catalysis, 1988, \textbf{109}, 367;
\doi{10.1016/0021-9517(88)90219-9}.
%
\bibitem{11} Wu N., Zhao M., Zheng~J.G., Jiang~C., Myers~B., Li~S., Chyu~M., Mao~S.X.,
Nanotechnology, 2005, \textbf{16}, 2878;
\doi{10.1088/0957-4484/16/12/024}.
%
\bibitem{12} Giamello E., Fubini B., Lauro P., Appl. Catal., 1986, \textbf{21}, 133;
\doi{10.1016/S0166-9834(00)81334-2}.
%
\bibitem{13} Losee D.B., A. Kassman I., Wilson P.A., J.~Catal., 1981, \textbf{67}, 226;
\doi{10.1016/0021-9517(81)90275-X}.
%
\bibitem{14} Nicula A., Stamires D., Turkevich J.,
 J.~Chem. Phys., 1965, \textbf{42}, 3684; \doi{10.1063/1.1695780}.
%
\bibitem{15} Yablokov Y.V., Ivanova T.A., Coord. Chem. Rev., 1999, \textbf{190-192},
1255;\\
\doi{10.1016/S0010-8545(99)00173-3}.
%
\bibitem{16}  Wei W.H., Wu S.Y., Dong H.N., Z.~Naturforsch. A, 2005, \textbf{60}, 541.
%
\bibitem{17} Newman D.J., B. Ng, Rep. Prog. Phys., 1999, \textbf{52}, 699;
\doi{10.1088/0034-4885/52/6/002}.
%
\bibitem{18} Yu W.L., Zhang X.M., Yang L.X., Zen B.Q., Phys. Rev. B,
 1994, \textbf{50}, 6756; \\ \doi{10.1103/PhysRevB.50.6756}.
 %
\bibitem{19} Newman J.D., Pryce D.C., Runciman W.A., Am. Mineral., 1978, \textbf{63}, 1278.
%
\bibitem{20} Petrosyan A.K., Khachatryas R.M., Sharoyas K.G., Phys. Stat. Sol. B,
 1984, \textbf{122}, 725; \doi{10.1002/pssb.2221220238}.
 %
\bibitem{21} Chakravarty A.S., Introduction to the Magnetic Properties of Solids.
Wiley-Interscience Publication, New York, 1980.
%
\bibitem{22} Griffith J.S., The Theory of Transition-Metal Ions.
Cambridge University Press, London, 1964.
%
\bibitem{23} McGarvey B.R., J. Chem. Phys., 1967, \textbf{71}, 51; \doi{10.1021/j100860a007}.
%
\bibitem{24} Gourier D., Colle L., Lejus~A.M., Vivien~D., Moncorge~R.,%
J.~Appl. Phys., 1988, \textbf{63}, 1144; \doi{10.1063/1.341139}.
%
\bibitem{25} Hodgson E.K., Fridovich I., Biochem. Biophys. Res. Commun.,
 1973, \textbf{54}, 270; \\ \doi{10.1016/0006-291X(73)90918-2}.
\end{thebibliography}
\end{document}